\documentclass[12pt]{article}
\pdfoutput=1 
\usepackage[footnotesize]{caption} 
\usepackage{amssymb}
\usepackage{epsfig}
\usepackage{amsmath}
\usepackage{booktabs}
\usepackage{graphicx}
\usepackage{graphics}

\usepackage{lscape}
\usepackage{float}
\usepackage{subfigure}
\usepackage{multirow}
\usepackage[toc,page]{appendix}
\usepackage{setspace}
\usepackage{color}
\usepackage{subfloat}
\usepackage{algorithm}
\usepackage{algpseudocode}
\newcommand{\bm}[1]{\mbox{\boldmath $#1$}}
\newcommand{\be}{\begin{equation}}
\newcommand{\ee}{\end{equation}}

\newcommand{\wh}{\widehat}

\newtheorem{theorem}{Theorem}

\begin{document}
\title{Analysis and Optimal Targets Setup of a Multihead Weighing Machine}
\author{
{\small Enrique del Castillo\footnote{Corresponding author. Dr. Castillo is Distinguished Professor of Industrial \& Manufacturing Engineering and Professor of Statistics. e-mail: exd13@psu.edu}}\\
{\small Department of Industrial and Manufacturing Engineering}\\
{\small The Pennsylvania State University, University Park, PA 16802, USA}\vspace{0.5cm}\\
{\small Alessia Beretta\footnote{Ms. Beretta is a Ph.D. student in Mechanical Engineering.} $\quad$ and $\quad$ Quirico Semeraro\footnote{Dr. Semeraro is a Professor in the Mechanical Engineering Department.}}\\
{\small
Dipartimento di Meccanica}\\{\small Politecnico di Milano, 20133 Milano, Italy}}

\date{\footnotesize \today}
\maketitle
\begin{abstract}
Multihead weighing machines (MWMs) are ubiquitous in industry for fast and accurate packaging of a wide variety of foods and vegetables, small hardware items and office supplies. A MWM consists of a system of multiple hoppers that are filled with product which when discharged through a funnel fills a package to a desired weight. Operating this machine requires first to specify the product weight targets or setpoints that each hopper should contain on average in each cycle, which do not need to be identical. The selection of these setpoints has a major impact on the performance of a MWM. Each cycle, the machine fills a package running a built-in knapsack algorithm that opens --or leaves shut-- different combinations of hoppers releasing their content such that the total weight of each package is near to its target, minimizing the amount of product ``given away". In this paper, we address the practical open problem for industry of how to determine the setpoint weights for each of the hoppers given a desired total package weight for a widely used type of MWM. An order statistic formulation based on a characterization of near-optimal solutions is presented. This is shown to be computationally intractable, and a faster heuristic that utilizes a lower bound approximation of the expected smallest order statistic is proposed instead. 
The setup solutions obtained with the proposed methods can result in substantial savings for MWM users. Alternatively, the analysis presented could be used by management to justify the acquisition of new MWM machines.
\end{abstract}

\baselineskip=14pt 

\section{Introduction}
A multihead weighing machine (hereafter a MWM, sometimes called a combinatorial weighing machine) is a computer-controlled machine used to fill a package with small products or parts with a given target weight. This machine has a wide range of applications in the food industry for packaging pasta, coffee beans, cereals, snacks, candies, vegetables, and even for packing poultry pieces and beef. Its applications cover also the packaging of non-food items, for instance, clips, nails, screws and a variety of other small hardware items.  
 Among the multihead weigher manufacturers, the one with the world leading position has 31,000 MWMs installed all over the world \cite{Ishida}. Despite their widespread use, analytical studies aimed at optimally setting up a MWM, a critical step affecting the performance of these machines, are lacking.
In this paper, we model and analyze a MWM and propose methods for its optimal setup.

\begin{figure}[htbp]
\begin{center}
    \includegraphics[width=0.6\textwidth]{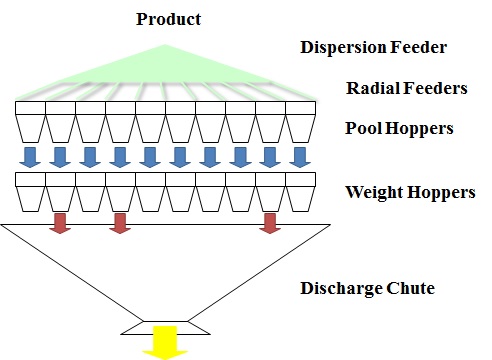}
    \caption{A single-layered multihead weighing machine.}\label{weigher}
\end{center}
\end{figure}

A MWM is composed of a system of feeders, a set of $H$ pool hoppers, a set of $H$ weight hoppers and a discharge chute to the packaging machine (Figure~\ref{weigher}). The product is continuously fed via a central dispersion feeder (usually a vibrating cone) and $H$ radial feeders (vibrating channels) to the pool hoppers. The role of the pool hoppers is to stabilize the product before dropping it into the weight hoppers. The average weight of product $\mu_i$, $i=1,...,H$, that each hopper should contain must be specified by an operator before starting the machine. 
 These average weights need not be identical.
Once the machine is started, each cycle, a built-in knapsack-like algorithm selects a subset of hoppers whose sum of observed weights is closest to the target value after which a computer opens the selected hoppers releasing the product through the discharge chute into the package. Some hoppers can therefore remain shut filled with product from cycle to cycle. One cycle is repeated for each package. The performance of a MWM heavily depends on the initial hopper weights $\{\mu_i\}$.
In industrial practice, operators currently use trial and error rules to setup the hopper weights based on the product to pack and the target weight of the package,  but such setting-up operation may be far from optimal.  In this paper, we focus on the analysis and optimal setup of MWMs with a single layer of hoopers (Figure \ref{weigher}), the most common type of MWM in industrial use.


Practically all of the extant technical literature related to MWMs (see, e.g., \cite{Kameoka,Karuno,Karuno2,Imahori}), which originates in Japan where MWMs were first developed, deals with the {\em on-line} problem of finding the best combination of hoppers to open in each cycle, proposing different versions of Knapsack formulations, but does not address the {\em off-line} problem of selecting the hopper weights {\em before} startup. The MWM we define below is somewhat related to canning problems \cite{Arcelus,Pollock} but they differ in that they deal with a single target filling setting problem, and more importantly, there is no selection combination problem involved.

MWM's are based on an empirically observed ``variance reduction" technique: it was noted that by filling a package from the combination of product from several hoppers, negative correlations are induced between the weights of product in opened hoppers given that they are random variables that are selected in each cycle subject to a constraint in their sum (which gives the package total weight). The negative correlations reduce the mean square error of the packages weight, ``giving away" less product while satisfying the target constraint.

The rest of the paper is organized as follows. The next section presents a mathematical formulation of the MWM setup problem and an exact approach for simple problems (i.e. when only few combinations of hoppers opening are considered). Next, the behavior of good solutions obtained by numerical search is characterized. These characteristics are then used in section 4 to develop a heuristic approach to the optimal machine setup. The paper ends with recommendations and directions for further research.

\section{Formulation of the optimal setup problem of a multihead weigher machine}
\label{section1}
Let $w_j$ be the observed weight of the product contained in the jth hopper in a particular cycle of operation, $j=1,2,...,H$ where $H$ is the number of hoppers in the machine. Each cycle the machine fills up a package with product released from a subset of the hoppers and the depleted hoppers are refilled. Assume $w_j$ is a realization of the random weight $W_j \sim N(\mu_j, \sigma_j^2=\alpha^2 \mu_j^2), \; \mu_j>0, j=1,2,...,H$ and assume each weight is independent of other weights $W_i (i \neq j)$. The proportionality constant $\alpha$ (with $\alpha<1$) is assumed known and given as it depends on the product to be packed. 
This relation between mean and variance of the weights is known to exist in this type of machines (e.g., see \cite{Kameoka}) and we have also observed it empirically. A {\em setup} of the machine consists of specifying the values of the setpoints $\bm \mu'=(\mu_1,\mu_2,...,\mu_H)$, to which, according to our assumption, also determine the hopper weight variances $\sigma_j^2, j=1,...,H$, for a given target value $T$ that specifies  the minimum weight content of each package to be filled. Once the machine is setup, the combinatorial weigher machine starts to fill packages of product, solving a knapsack algorithm per package. Our goal is to determine the best setpoints $\bm \mu$ according to some specific criteria on the weight content of the packages.

\begin{figure}[htbp]
\centering
\resizebox{12cm}{5.0cm}{\includegraphics{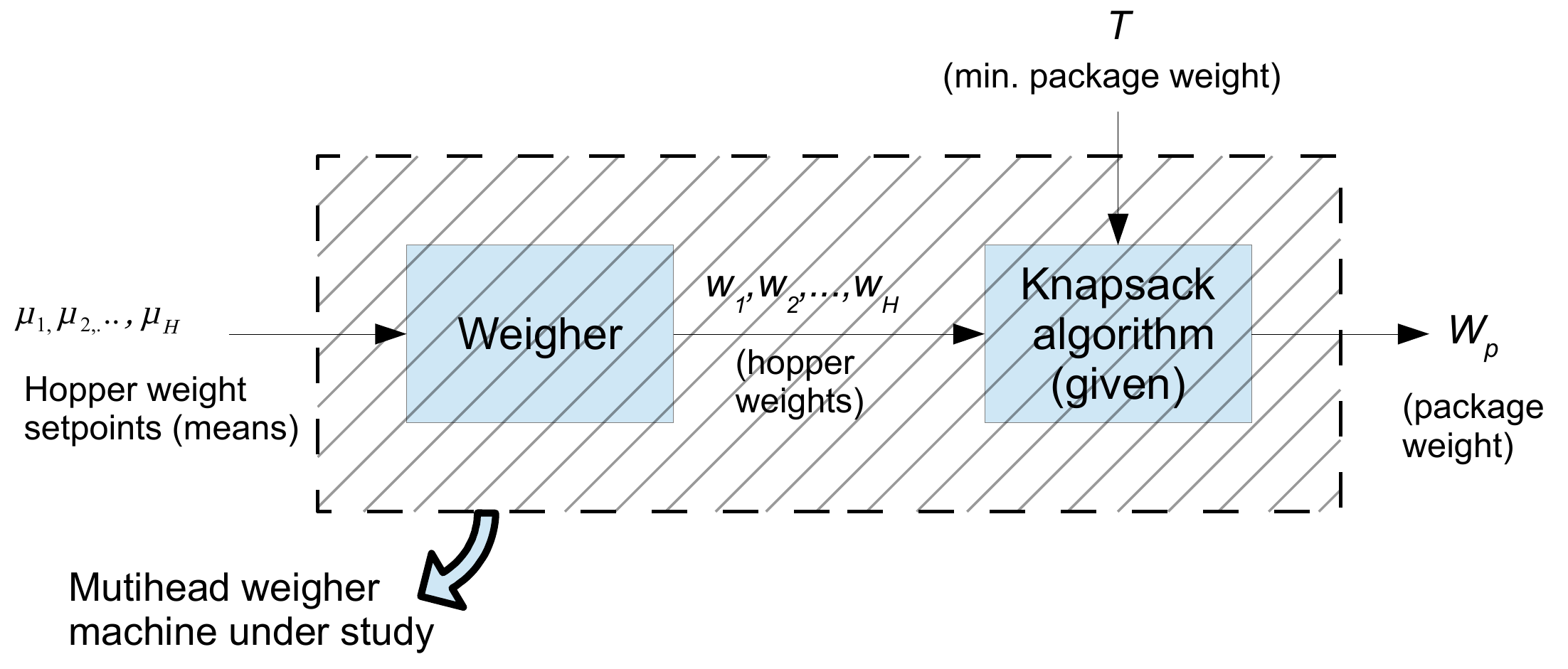}}
\caption{Information flow in the system under study. The setup problem in a multihead  weigher machine consists in finding values of the hopper weights $\mu_1,...,\mu_H$ that optimize some property of the resulting package weight $w_p$. The built-in knapsack algorithm the machine comes equipped with is considered internal and given. Therefore, the system under study (delimited by dashed lines and hatched) requires determination of the ``inputs" (the $\mu_i$'s) to optimize some property of the ``output" $w_p$. \label{figMachineDiagram}}
\end{figure}

While there are different  knapsack formulations that have been reported in the MWM literature, most of them utilize a linear objective function and linear constraints. In this section, we assume the machine has a built-in algorithm that solves for each package the deterministic knapsack problem:
\be
\label{knapsack}
\min \quad w_p=\sum_{j=1}^H \delta_j w_j \;\;\;\; \mbox{ subject to: } \;\;\;\;  w_p=\sum_{j=1}^H \delta_j w_j>T
\ee
where $\delta_j$ is either 0 or 1. In this formulation, the total package weight $w_p$ is required to be as small as possible but larger than the given target package weight $T$. Prior to observing the hopper weights $\{W_j=w_j\}$ in any cycle, the total package weight $W_p$ is the minimum of $K$ dependent, not identical normal random variables $X_i$ for $i=1,2,...,K$, subject to the constraint $W_p>T$, where $K$ equals the total number of possible combinations of opened/closed hoppers from which the knapsack algorithm can select  (choosing the $\delta_j$ variables above). 
We hasten to point out that {\em we are not concerned with solving the knapsack problem; the knapsack problem is internal to the machine and considered given}. 
We are concerned with determining the setpoints of the machine, i.e., the mean weights in each hopper, which are the ``inputs" of the system as depicted in Figure \ref{figMachineDiagram}.

The optimal setpoints could be found from the distribution of the optimum objective function value (i.e., the package weight $W_p$) of the random Knapsack (\ref{knapsack}). However, there are only limited results related to this distribution (\cite{Prekopa}, p. 526). They are asymptotic results as the number of hoppers $H \rightarrow \infty$ under the assumption the hopper weights $W_i$'s are $U(0,1)$ random variables, which is clearly not our case. In the remainder of this section we describe how to compute the exact {\em moments} of the optimal weight package $W_p$ in problem (\ref{knapsack}) and how this leads very rapidly to computational complexities in practice.

If all possible combinations of any number of hoppers can be selected to open (or close) in a cycle, then clearly there are $K=2^H-1$ combinations, since we can assume that at least one hopper will open in each cycle to let some product get in the package.  From now on, we use the term {\em combination} to refer to a specific selection of $\delta_j, \; j=1,2,...,H$ variables in (\ref{knapsack}), i.e., to a specific selection of hoppers that are opened in a cycle. We also apply this term by extension to the package weight $X_i=\sum_{j=1}^H \delta_j W_j$ that combination $i$ generates ($i=1,...,K$). Hence we refer to $K$ as the number of possible combinations.

To consider the different combinations of weights, let matrix $\bm P =\{p_{ij}\}$ be a $K \times H$ matrix where $\{p_{ij}=\delta_j|i\}$, with $p_{ij}=1$ if combination $i$ includes opening hopper $j$ and $p_{ij}=0$ if otherwise, $i=1,...,K$. We then define the random vector of combinations $\bm X'=(X_1,X_2,...,X_K)$ as:
\be
\bm X = \bm P \bm W
\ee
where $\bm W'=(W_1,W_2,...,W_H)$ follows a multivariate N$(\bm \mu, \bm \Sigma_W)$ distribution with $\Sigma_W = \mbox{diag}(\alpha^2 \mu_j^2)$. It follows that for the different weight combinations that can be formed we have that:
\[
\bm X \sim N(\bm \Theta, \bm \Sigma) \quad \quad \mbox{where} \quad \quad \bm \Theta= E[\bm X] = \bm P \bm \mu, \; \; \mbox { and } \; \; \bm \Sigma = \mbox{Var}(\bm X) = \bm P \bm \Sigma_W \bm P'.
\]
The $K \times K$ matrix $\bm \Sigma$ includes the covariances between the random weights resulting from the different combinations of selected hoppers, some of which may be large, depending on the $K$ combinations to consider. If combinations that ``share" many hoppers are included, $\bm \Sigma$ may be close to rank deficient.

{\bf Problem definition. } The optimal off-line MWM setup problem we address, that corresponds to the on-line knapsack problem (\ref{knapsack})  requires solving:
\[
\min_{\bm \mu} \;\; \mbox{MSE}(W_p | W_p > T)
\]
that is, finding the hopper setpoints such that the mean square error of the package weight $W_p=\min(X_1,X_2,...,X_K)$ with  $W_p>T$ is minimized, since it is desired that no package should weigh less than $T$. The distribution of $W_p$ is a function of the hopper mean weights $\bm \mu$. Note that the $\mu_j$ are not required to be integers. Therefore, we must first {\em find the distribution of a constrained smallest order statistic of a set of correlated normal variables}, and setup an optimization problem with it. As far as we know, there are no published results related to such problem in the Order Statistics literature. Afonja \cite{Afonja} found expressions for the first two moments of the {\em unconstrained} maximum order statistic of a set of correlated normals. The constraint $W_p>T$ considerably increases the computational complexity when obtaining the moments in exact form.

We show next that the computational expense of the expressions needed to obtain the moments of such constrained minimum is too large in general, motivating the approximate approach shown in a later section.

\begin{figure}
\begin{center}
\resizebox{13cm}{9.0cm}{\includegraphics{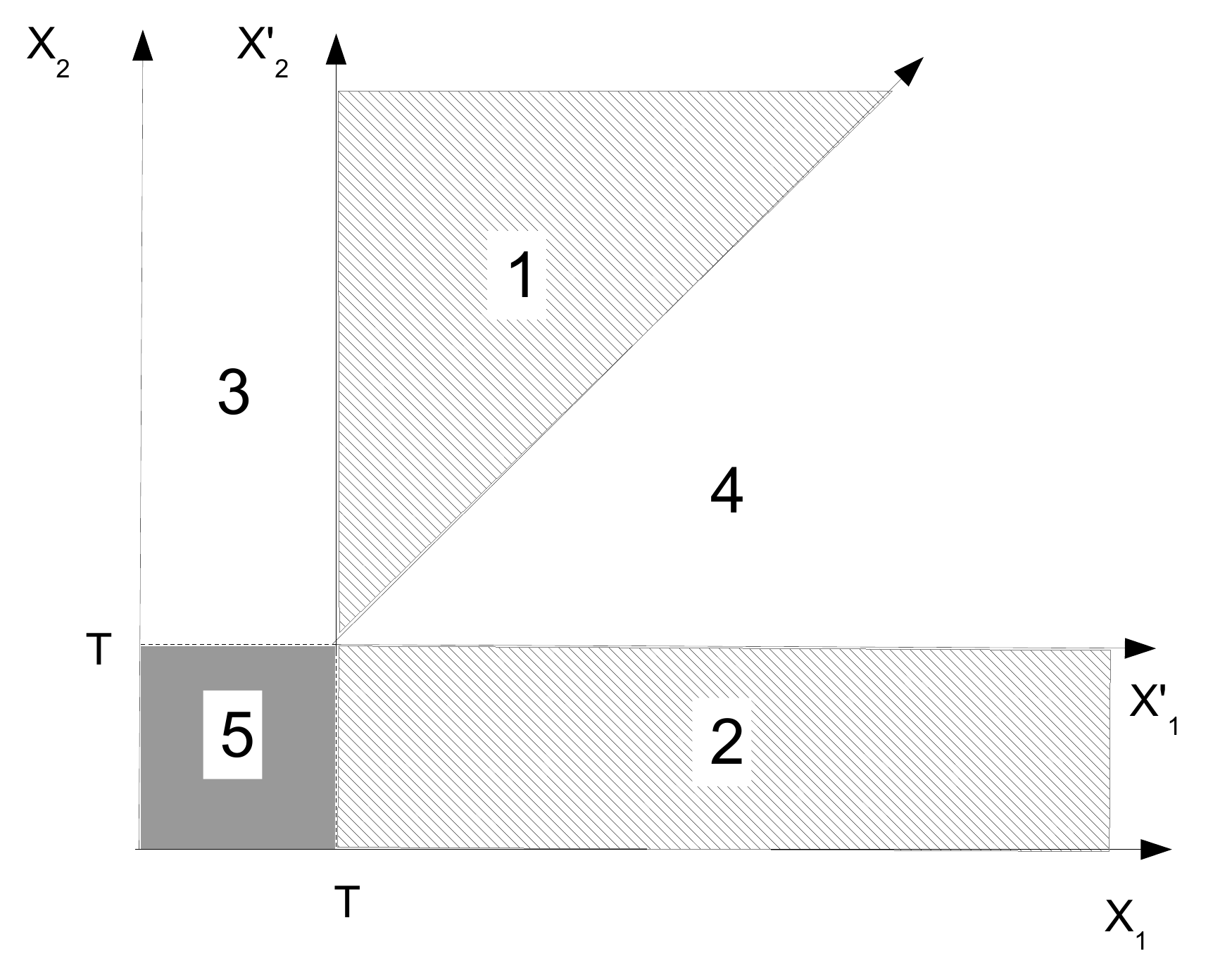}}
\caption{A ``toy" weigher machine setup problem with two combinations. Hashed regions 1 and 2 are areas where combination $X_1$ is better than combination $X_2$ ($X_1$ is closer to and greater than $T$); in regions 3 and 4 $X_2$ is better ($X_2$ is closer to and greater than $T$). Shaded region 5 is infeasible. The search for a feasible solution needs to be undertaken over regions 1 through 4.\label{fig1}}
\end{center}
\end{figure}

Consider first the simple case where there are only two combinations of hoppers ($K=2$). Figure \ref{fig1} shows the regions where each random variable $X_i \; (i=1,2)$ achieves the minimum weight. The two hashed regions (numbers 1 and 2) are areas over which $X_1$ is minimum and greater than $T$; symmetrically, the two unshaded areas (3 and 4) are where combination $X_2$ is the minimum and greater than $T$.
We point out that we do not seek to find $\min(X_1|X_1>T, X_2|X_2>T)$, which would correspond only to taking the minimum over areas 1 and 4 in Figure \ref{fig1}.  This is not what we seek, since, for instance, we could have $X_1>T$ but $X_2<T$ and still have found a feasible solution to our packing problem; all we need is at least one combination to be larger than $T$. Thus, for $H=2$ we need to search in all four areas 1 to 4 in Figure \ref{fig1}. If $W_p=\min(X_1,X_2)$, what we seek is $W_p|W_p>T$, i.e., {\em the minimum of the combinations is what is constrained, not the individual combination weights}.

Define $X_i'=X_i-T$, as in the figure. The weigher machine must select one of the variables $X_i'$ to fill up a package. The rth moment  of the selected weight $W_p$ is therefore given by:
\begin{eqnarray}
E[W_p^r|W_p>T] &=& \int_{0}^{\infty} x_1'^r \int_{x_1'}^{\infty} \bm \phi(\bm X') \; dx_1 dx_2
+\int_{0}^{\infty} x_1'^r \int_{-T}^{0} \bm \phi(\bm X') \; dx_1' dx_2' \nonumber \\
&+& \int_{0}^{\infty}  x_2'^r \int_{x_2'}^{\infty}  \bm \phi(\bm X') \; dx_2' dx_1'
+ \int_{0}^{\infty}  x_2'^r \int_{-T}^{0} \bm \phi(\bm X) \; dx_2' dx_1' \;\;\;\;\;\;\;\;\;\;
\end{eqnarray}
where $\bm \phi(\bm X')$ is the (bivariate) normal density of $\bm X'$, i.e., N$(\bm \Theta-\bm T, \bm \Sigma)$ with $\bm T$ equal to a $K$-vector filled with the package target weight $T$. The four terms correspond to integrals over the probability measure in areas 1, 2, 4 and 3 in Figure \ref{fig1}, respectively. Solving the setup problem for a combinatorial machine in this case consists in minimizing  MSE$(W_p|W_p>T)=\mbox{Var}(W_p|W_p>T)+(E[W_p|W_p>T]-T)^2=E[W_p^2|W_p>T]-E[W_p|W_p>T]^2+(E[W_p|W_p>T]-T)^2$ with respect to $\mu_1$ and $\mu_2$. Therefore, the integrals above will need to be performed several times inside an optimization routine.

Consider next the case of $K=3$ different combinations the MWM can select from. In this case we will get more rectangular areas similar to regions 2 and 3 in Figure \ref{fig1}. Specifically, the rth moment of the selected weight is given by the expression:
\begin{eqnarray}
E[W_p^r | W_p > T] &=& \sum_{i=1}^3 \left\{ \int_{0}^{\infty} x_i'^r \int_{x_i'}^{\infty} \int_{x_i'}^{\infty} \bm \phi(\bm X') \; d\bm x'
+\int_{0}^{\infty} x_i'^r \int_{-T}^{0} \int_{-T}^{0} \bm \phi(\bm X') \; d\bm x' \right. \nonumber \\
&+& \left. \int_{0}^{\infty} x_i'^r \int_{x_i'}^{\infty} \int_{-T}^{0} \bm \phi(\bm X') \; d\bm x'
+\int_{0}^{\infty} x_i'^r \int_{-T}^{0} \int_{x_i}^{\infty} \bm \phi(\bm X') \; d\bm x' \right\} \;\;\;\;\;\;\;\;\;\;
\end{eqnarray}
which is an expression with twelve 3-dimensional integrals where
\[ 12 =\left[\left(\begin{array}{c}K-1\\K-1\end{array} \right)+\left(\begin{array}{c}K-1\\K-2\end{array} \right) + \cdots + \left(\begin{array}{c}K-1\\0\end{array} \right)\right] \cdot K=2^{K-1} \cdot K
\]  for $K=3$.

Evidently,  $2^{K-1} \cdot K$ grows very fast. If {\em all} the combinations of $H$ hoppers are considered (so $K=2^H-1$) the total number of $2^H-1$-dimensional integrals, $2^{2^H-2} \cdot (2^H-1)$, grows extraordinarily fast, see Table \ref{tab1}.

\begin{table}[htdp]
\begin{center}
\begin{tabular}{crc}
{No. of hoppers ($H$)}&{No. of Integrals}&{Dimension of each integral ($K$)}\\
\hline
1&1&1\\
2&12&3\\
3&448&7\\
4&245760&15\\
5&3.3286 e10 & 31\\
8&7.3817 e78 & 255\\
\end{tabular}
\caption{Number of multidimensional integrals needed to compute a moment $E[W_p^r|W_p>T]$ of the package weight assuming all $2^H-1$ possible combinations are considered. \label{tab1}}
\end{center} 
\end{table}%

Although in practice not all possible combinations need to be considered, the exponentially increasing number of combinations and the required computations  render an exact approach intractable for problems of realistic size. Therefore, we proceed next to develop approximate approaches to the solution of this problem. We first characterize the properties of near-optimal solutions obtained by a search procedure in order to seek similarly good solutions in a heuristic approach to be discussed in section \ref{section3}. 

\section{Characterization of near optimal solutions to the MWM  setup problem}

As mentioned earlier, hereon by a `good' solution we mean a low (or near minimum) MSE solution subject to the constraint $W_p > T$. For computational simplicity, in this section we only consider simulated MWMs with $H=4$ and $H=5$ hoppers when none, one, or two hoppers can remain shut in a cycle. The target value $T=500$ and the parameter $\alpha=0.123$ were used throughout. There is no loss of generality since the characteristics discussed in this section are independent of these parameters.
 \begin{figure}[htbp]
\begin{center}
\resizebox{16cm}{7cm}{\includegraphics{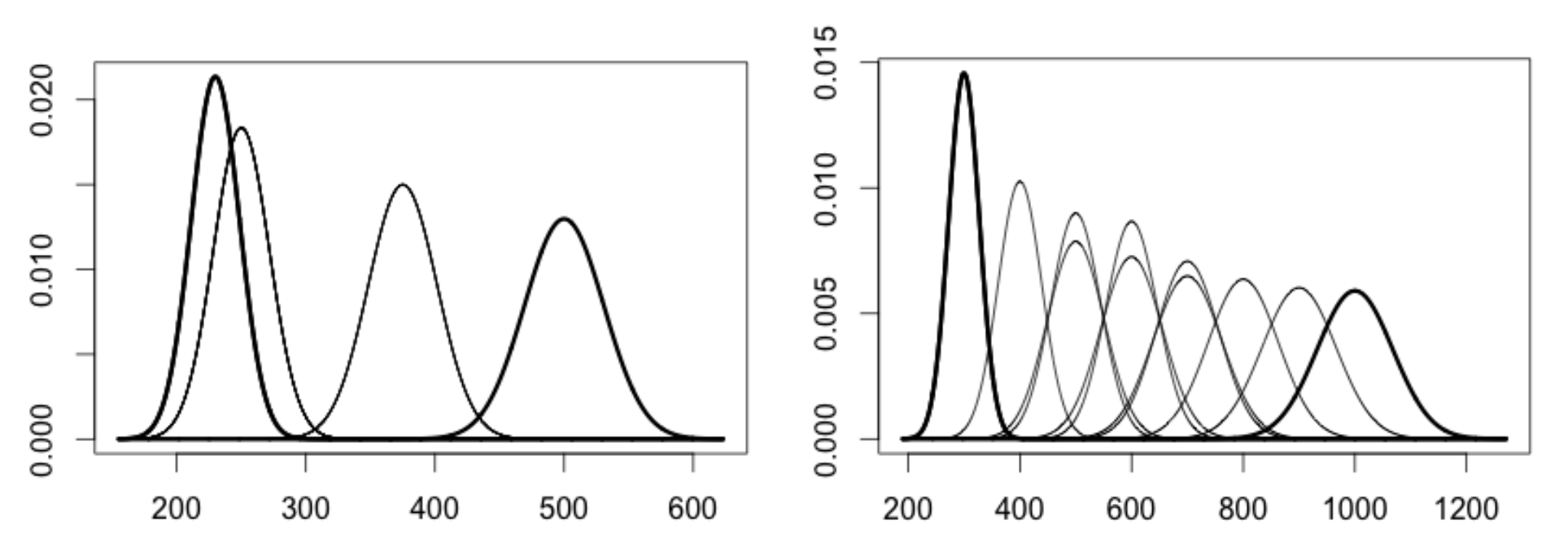}}
\caption{Marginal density plots of the  combination weights $\{X_i\}$ generated by two non-optimal solutions of a weigher problem with $H=4$ hoppers. Plot on the left is for the solution $\bm \mu'=(500/4,500/4,500/4,500/4)$. Plot on the right is for the solution $\bm \mu'=(400,300,200,100)$. Combinations up to two closed hoppers were considered. Darker density lines correspond to $X_{[1]}$ and $X_{[K]}$.\label{fig21}}
\end{center}
\end{figure}
Figure \ref{fig21} shows the marginal densities of the $K=11$ combinations $X_i, i=1...,11$, generated in a $H=4$ hopper problem for two typical, but not optimal solutions.  The marginal densities of the first order statistic $X_{[1]}$ and the last order statistic $X_{[K]}$ (the moments of these densities were obtained as described in Appendix A) are highlighted. The plot on the left corresponds to the setpoints $\bm \mu'=(T/4,T/4,T/4,T/4)$, a "logical" solution in which all hopper setpoint weights are equal. As it can be seen, the weigher machine will have only one combination available to fill the $T=500$ gr. packages, namely, the density of the largest order statistic which corresponds to all hoppers opened in each cycle. Note also how there are many densities that overlap perfectly, since there are $K=11$ combinations, but only three are uniquely different densities. This is a very poor solution, since there will be about a 50 \% chance that the constraint $W_p>T=500$ will not be satisfied. The graph on the right shows the marginal densities for a different `ladder' solution with setpoints $\bm \mu'=(400,300,200,100)$. The marginal densities are much more dispersed, which is better than the previous solution. But note how the densities are not dispersed symmetrically around $T=500$, and in particular, note how
there are only a few `good' combinations, i.e., densities located near or above $T=500$, so the performance will be far from optimal.

\begin{figure}[htbp]
\begin{center}
\resizebox{16cm}{7cm}{\includegraphics{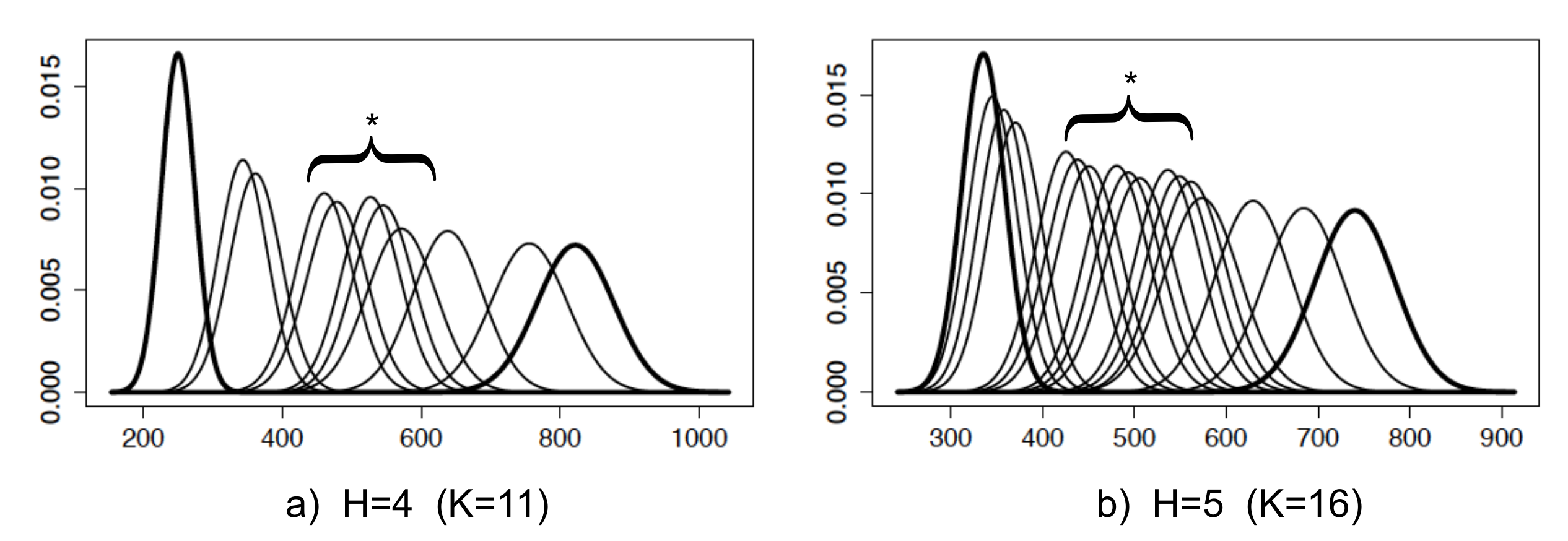}}
\caption{Density plots of the near-optimal $\{X_i\}$ combination weights for a problem with a) 4 hoppers and b) 5 hoppers. Combinations up to two closed hoppers were considered. Darker density lines correspond to $X_{[1]}$ and $X_{[K]}$. Note the cluster of densities (*) around $T=500$ (see text).\label{fig2}}
\end{center}
\end{figure}

In contrast with the two previous solutions which can be classified as poor, Figure \ref{fig2} shows the marginal densities for near-optimal solutions for the $H=4$ case and the $H=5$ cases, obtained  by a simple search of the best combination as reported in \cite{Beretta}. This search attempts to minimize the MSE of the selected weights $W_p$ such that $W_p>T$. The solution found for the $H=4$ problem is $\bm \mu'=(294.5,276.7,183.7,66.6)$. For the $H=5$ hopper MWM there are $K=16$ combinations possible (with up to 3 hoppers remaining shut), and the densities shown in the figure correspond to the solution $\bm \mu'=(203.7,178.6,110.9,191.0,55.7)$.
Some useful characteristics of the solutions $\bm \mu^*$ that generated the combinations shown in Figure \ref{fig2} are shown in Table \ref{tab2}.
\begin{table}[htdp]
\begin{center}
\begin{tabular}{c|c|c}
Property&{$H=4$}&{$H=5$}\\
\hline
 $\bm \mu^*$&(294.9, 276.7, 183.7, 66.6) & (203.7, 191.0, 178.6, 110.9, 55.7)\\
$(E[X_{[1]}]+E[X_{[K]}])/2$& 536.0 & 537.7\\
$\overline{\bm \Theta}$&523.2&508.7\\
-log(det$(\bm \Sigma)$)&49.67&89.42\\
$p \equiv \sum_{i=1}^K P\{X_i \in (T,T+100)\}$&2.78&4.79\\
$c(\bm \Theta, T)$&4&7\\
-log(det($\bm M$))&26.10&66.19\\
\hline
$\wh{E}(W_p|W_p>T)$&522.8 &514.2\\
$\wh{\mbox{Var}}(W_p|W_p>T)$&301.1 &129.2\\
$\wh{\mbox{MSE}}(W_p|W_p>T)$&822.7 &331.1\\
\hline
\end{tabular}
\caption{Some characteristics of the near optimal (min MSE) solutions for $H=4$ and $H=5$ hoppers, $T=500, \alpha=0.123$ obtained by searching combinations with up to 2 hoppers shut.  A heuristic method is developed to find solutions that mimic all these characteristics. Quantities listed below the line were estimated via simulation for the solution in question.\label{tab2}}
\end{center}
\end{table}

An interesting characteristic common to all near-optimal both solutions such as those shown in Table \ref{tab2} is that the average of the two extreme order statistics, $(E[X_{[1]}]+E[X_{[K]}])/2$ remains above $T$. At the same time, the combinations are quite dispersed. This leads to our first empirical characterization of a good solution.

{\em Characteristic 1:} in a good solution, the combinations $\{X_i\}_{i=1}^K$ should be such that the average of the two (unconstrained) extreme order statistics is larger than $T$.

Likewise, good solutions such as those in Table \ref{tab2} have a cluster of {\em many} densities around $T$, indicated with a ``*" in Figure \ref{fig2}. These densities should not all be near identical, but should differ, thus some variability in them is desirable. We hence define $u(\bm \Theta, T)$ to be the vector of values $\Theta_i$ that differ by at least $0.04 T$ of other $\Theta_i$'s:  $u(\bm \Theta, T) \equiv {\tt unique}(\bm \Theta - (\bm \Theta \mod 0.04 \cdot T))$, where we assume we have available a function  {\tt unique} that returns the different items in a vector. Then, we define the number of densities with significantly different locations in an interval around $T$  $(0.8 \cdot T, 1.2 \cdot T)$ as:
\[
c(\bm \Theta, T) = \sum_{i} \{(u_i > 0.8 \cdot T) \mbox{ and } (u_i < 1.2 \cdot T)\}
\]
These non-identical densities around $T$ provide significantly different combinations to the on-line knapsack algorithm to select from and fill up a package. Hence,  for a near-optimal solution we observe that the sum of the marginal probabilities $\sum_{i=1}^K P\{X_i \in (T,1.2 \cdot T)\}$ is large relative to a non-optimal solution. This can be summarized in:

{\em Characteristic 2:} good solutions generate combinations whose densities cluster around $T$ and are characterized by a large count number of uniquely different densities $c(\bm \Theta, T)$ as defined above. Thanks to this cluster of densities, good solutions are associated to a relatively large value of $p(\bm \Theta, T) \equiv \sum_{i=1}^K P\{X_i \in (T, 1.2 \cdot T)\}$, the probability of package weights just above $T$, compared to non-optimal solutions.

The marginal density functions of each combination, however, are usually positively correlated and the probability $p(\bm \Theta, T)$ is  not very informative by itself unless the correlations are accounted for. {\em A set of highly positively correlated combinations behaves essentially as fewer combinations}.
 The determinant of $\bm \Sigma$ is a simple measure of the global degree of correlation in the combinations. It was observed that the values of log(det$(\bm \Sigma))$ are relatively large for the near-optimal solutions compared to non-optimal solutions. This implies that in a good solution, the resulting combinations $\{X_i\}_{i=1}^K$ are relatively less correlated.  Thus, the densities  should be as little correlated as possible (especially those close to $T$), in order to provide as uncorrelated combinations as possible to the knapsack algorithm. Apart of the cluster of good combinations close to $T$, other densities should disperse in a symmetrical manner around $T$. To emphasize these two aspects, low correlation of the combinations located near $T$ and variability of their locations around $T$, we define the matrix:
\[
\bm M = \bm \Sigma + (\bm \Theta- \bm T)(\bm \Theta - \bm T)'
\]
which adds to $\bm \Sigma$ a measure of variability of the locations of the densities around $T$ in the second term. The values observed for log(det($\bm M$))  for the near-optimal solutions are large relative to non-optimal solutions (Table \ref{tab2}). This has the effect of dispersing the densities, avoiding too similar densities around $T$. In summary, we have the following.

{\em Characteristic 3.} A good solution provides large values of det$(\bm M) \equiv |\bm M|$  relative to non-optimal solutions.

Finally, we point out an obvious but useful property of any solution $\bm \mu$. The combinatorial weigher machine does not give preference to any hopper over the others, and each hopper can be set in exactly the same way as the others. In other words, we have:

{\em Characteristic 4.} The quality of a solution $(\mu_1,\mu_2,...,\mu_H)$, as measured by any function of $W_p$ (e.g. MSE($W_p|W_p>T$)) is invariant to any permutation $\sigma(\cdot)$ of the hoppers indices $(\mu_{\sigma(1)},\mu_{\sigma(2)},...,\mu_{\sigma(H)})$.

This is useful since we can reduce the search conducted by any optimization algorithm to the region $\mu_1>\mu_2>\cdots > \mu_K$.

\section{A heuristic optimization model for determining the MWM hopper setpoints}
\label{section3}

The four characteristics identified in the previous section as common to all near-optimal solutions can be used to devise a heuristic algorithm for setting up a MWM. We therefore propose to solve:
\be
\mbox{maximize} \; \quad \mbox{log}|\bm M| + p(\bm \Theta, T) + c(\bm \Theta, T)
\label{objective}
\ee
subject to:
\begin{eqnarray}
\frac{E[X_{[1]}]+E[X_{[K]}]}{2}&>&1.1  \cdot T \label{first}\\
\mu_1>\mu_2>\cdots &>& \mu_K \label{order}\\
0 < \mu_i & < & f \cdot T \label{fractionT}
\end{eqnarray}
The log used in (\ref{objective}) is a standard way of scaling (up) a numerically small determinant. The upper limit in (\ref{fractionT}) is aimed to reduce the search space of each setpoint. It was observed that as $H$ increases, the optimal values of the $\mu_i$ decrease, so increasingly smaller values of $f$ with $f<1$ should be used.

Finally, in addition to constraints (\ref{first})-(\ref{fractionT}) it is necessary to ensure that there will always be at least a feasible combination in a cycle, i.e., one combination, that in which all hoppers open, should generate more product than $T$ with very high probability. Under the assumptions discussed in section \ref{section1}, this can be expressed as:
\[
\Phi\left(\frac{T-\sum_{i=1}^H \mu_i}{\alpha \sqrt{\sum_{i=1}^H \mu_i^2}} \right) \leq \epsilon
\]
which, since $\mu_i>0$ can be written as:
\be
\label{feasibility}
||\bm \mu||_1 + \alpha ||\bm \mu||_2 \Phi^{-1}(\epsilon) \geq T
 \ee

 The complete heuristic solution consists in maximizing (\ref{objective}) subject to constraints (\ref{first})-(\ref{feasibility}). To solve it, we use an Augmented Lagrange routine in R's {\tt Nloptr} library (see Appendix B for details of our R implementation). Since the optimization problem is clearly non-convex, the optimizer was started from a grid of initial near-feasible trial points (equations (\ref{order}), (\ref{fractionT}) and (\ref{feasibility}) are satisfied at the initial points, although (\ref{first}) may not be satisfied). Table \ref{tab3} shows results for $H=4$ and $H=5$ when the same combinations and parameters as for the solutions in Table \ref{tab2} were used.

 \begin{table}[htdp]
\begin{center}
\begin{tabular}{c|c|c}
Property&{$H=4$}&{$H=5$}\\
\hline
 $\bm \mu^*$&(267.4, 259, 234.6, 57.7) & (228.1, 200.3, 161.9, 113.5, 61.0)\\
$(E[X_{[1]}]+E[X_{[K]}])/2$& 549.9 & 549.6\\
$\overline{\bm \Theta}$&521.0&525.8\\
-log(det$(\bm \Sigma)$)&49.79&89.07\\
$p \equiv \sum_{i=1}^K P\{X_i \in (T,T+100)\}$&3.62&4.79\\
$c(\bm \Theta, T)$&6&9\\
-log(det($\bm M$))&26.2&65.84\\
\hline
$\wh{E}(W_p|W_p>T)$&520.7 &514.6\\
$\wh{\mbox{Var}}(W_p|W_p>T)$&340.3 &132.6\\
$\wh{\mbox{MSE}}(W_p|W_p>T)$&769.8* &347.0\\
\hline
\end{tabular}
\caption{Solutions obtained with the heuristic method (\ref{objective}-\ref{feasibility}) for $H=4$  and $H=5$ hoppers, $T=500, \alpha=0.123$, $\epsilon=1e-05$, $f=0.6$ ($H=4$) and $f=0.5$ ($H=5$), 100 initial trials (up to 2 hoppers can remain shut).  Quantities listed below the line were estimated via simulation for the solution in question. Compare to Table \ref{tab2}  (* indicates the heuristic gives better MSE solution). The exact expression for $E[X_{[1]}]$ was used.\label{tab3}}
\end{center}
\end{table}

As can be seen, the solutions found with the heuristic approach are very close to  those in Table \ref{tab2}, found by a simple search. For $H=4$ the heuristic provides a slightly better MSE, whereas for $H=5$ the opposite occurs.  Although characterizing the solutions increases the understanding of the problem, the heuristic method in this section is only computationally feasible when the number of combinations $K$ considered is very small. We now present a simple modification of this heuristic which allows a user to tackle  larger problems.

\subsection{Modified optimization heuristic for faster solution}

The main computational bottleneck of the heuristic optimization is the {\em exact} computation of the expected values of the extreme values in constraint (\ref{first}). We observed that while $E[X_{[K]}]$ could be well approximated with max$(\Theta_i)$, a similar approximation is not possible for $E[X_{[1]}]$. As shown in Appendix A, computing this expectation requires $K$ K-dimensional Normal integrals, which can only be attempted for small $K$. Unfortunately, we found constraint (\ref{first}) to be critical and hence it cannot be removed.

As an alternative, we could use a fast to compute lower bound $LB(E[X_{[1]}])$ instead of the computationally expensive $E[X_{[1]}]$ in  constraint (\ref{first}). Bertismas et al. \cite{Bertsimas} (Theorem 4) give a  useful closed-form lower bound  for $E[X_{[1]}]$ in a collection of possibly correlated normals. This bound is given by:
\begin{eqnarray}
E[X_{[1]}] &\geq& LB(E[X_{[1]}]) \equiv -\frac{1}{2} \left[ \sum_{i=1}^K \left(- \mu_i + \sqrt{\left(-\mu_i - \max_i\left\{- \mu_i + \frac{K-2}{2 \sqrt{K-1}} \sigma_i \right\} \right)^2 + \sigma_i^2} \right) \right] \nonumber \\
& - & \frac{2-K}{2} \max \left\{ - \mu_i + \frac{K-2}{2 \sqrt{K-1}} \sigma_i \right\}
\label{LBmin}
\end{eqnarray}
Bound (\ref{LBmin}) is extremely fast to compute compared to the exact moment, as it requires no integration. Table \ref{tab4} shows some computation times and MSE values for comparison purposes when solving (\ref{objective}-\ref{feasibility}) using the exact moment and its lower bound approximation.

\begin{table}[htdp]
\begin{center}
\begin{small}
\begin{tabular}{rrrrrrr}
{}&{}&{Time using}&{Time using}&{$\overline{\mbox{MSE}}\; (\wh{\mbox{sd}}\mbox{(MSE)})$}&{$\overline{\mbox{MSE}}\; (\wh{\mbox{sd}}\mbox{(MSE)})$}&{P-value for}\\
{$H$}&{}&{ $LB(E[X_{[1]}])$}&{exact }&{using exact}&{using}&{$H_1:$ exact}\\
{(max shut)}&{$K$}&{ in (\ref{first})}&{$E[X_{[1]}]$ in (\ref{first})}&{$E[X_{[1]}]$}&{$LB(E[X_{[1]}])$}&{$E[X_{[1]}]$ better}\\
\hline
4(2)	&	11	&	5.2	&	457.5	&	772.2 (7.32)	&	800.4 (5.78)	&	0.0000	       \\
4(3)	&	15	&	6.0	&	2001.3	&	701.3 (10.06)	&	672.6 (18.10)	&	1.0000	\\
5(2)	&	16	&	6.1	&	3167.2	&	349.8 (1.54)	&	391.8 (2.28)	&	0.0000       \\
\end{tabular}
\end{small}
\end{center}
\caption{Computing times (in seconds) for solving the setup weigher problem (\ref{first})-(\ref{feasibility}) using the exact or lower bound for the smallest order statistic $E[X_{[1]}]$ for problems with few combinations $K$.  In all cases, 100 initial trials, $\alpha=0.123$, average and standard errors of MSEs estimated based on 50,000 simulated cycles. Times on an Intel 2.93 GHz Core 2 PC running R. \label{tab4}}
\end{table}%
We have coded both versions of the heuristic, using the exact $E[X_{[1]}]$ as described in Appendix A and the lower bound (\ref{LBmin}) using R (see Appendix B). 
The computing time depends on $K$, the total number of combinations generated. Table \ref{tab4} shows the performance of solutions obtained with the exact moment only up to $K =16$ given the high computational times due to the multidimensional integrals involved. In contrast, it is notable how the computing time required for finding solutions using the moment lower bound scales  well with $K$ (Table \ref{tab5}).  For the three cases where we are able to compare the performance of the heuristic with the exact moment and with its lower bound approximation, the differences in MSE are inconclusive about which one is better overall. The exact moment provided an statistical better MSE solution in the cases $H=4$ (up to 2 shut) and $H=5$ (up to 2 shut), but the lower bound approximation provides a better MSE solution in the case $H=4$ (up to 3 shut).

As the number of hoppers increases and one considers combinations where more hoppers remain shut, a numerical problem occurs: $\bm \Sigma$ and hence $\bm M$ become very ill-conditioned, up to a point when $det(\bm \Sigma)$ is numerically zero. In our experiments this occurred (for $\alpha=0.123$) when $H>10$ and all $comb(H,3)$ were considered. This ill-conditioning can be reduced if the combination that consists in all hoppers opening is not considered. This was implemented in our computer code (Appendix B), which permitted us to solve problems for $H>10$.
\begin{table}
\begin{center}
\begin{small}
\begin{tabular}{cc|ccc}
{$H$}&{$K$}&\multicolumn{3}{|c}{Lower bound heuristic}\\
{(max shut)}&{(\# combs.)}&{Time}&{$\overline{\mbox{MSE}}$}&{$\wh{\mbox{sd}}(\mbox{MSE})$}\\
\hline
6(2)	&	22	&	6.2	    &		133.8 &1.58	
    \\
7(2)	&	29	&	7.7	    &		102.4 &0.80	
       \\
8(2)	&	37	&	10.5	&		80.5&0.92	
   \\
6(3)	&	42	&	10.7	&		100.4&0.88	
    \\
9(2)	&	46	&	16.9	&		111.9&1.20	    \\
7(3)	&	64	&	13.9	&		34.3&0.68	
  \\
12(2)   &	79	&	17.1	&		12.8&0.27	
    \\
8(3)	&	94	&	20.9	&		23.5&1.24	\\
9(3)	&	130	&	36.7	&		8.98&0.34		
  \\
12(3)   &	299	&	162.1   &		1.23&0.02		
\end{tabular}
\end{small}
\end{center}
\caption{Computing times (in seconds) for obtaining a solution to the setup weigher problem and the resulting average and estimated standard error of the MSE values over 10 simulations  for larger values of the number of combinations ($K$). Times and MSEs comparisons for solutions obtained solving (\ref{first})-(\ref{feasibility}) with the lower bound $LB(E[X_{[1]}])$ approximation in (\ref{first}) (using $f=0.3$ for all cases in (\ref{fractionT}) and $\alpha=0.123$). MSEs estimated based on 50,000 simulated cycles (if $K\leq 50$) or 10,000 cycles (if $K \geq 50$). Times on an Intel 2.93 GHz Core 2 PC running R. \label{tab5}}
\end{table}

\subsection{Discussion: Recommendations for setting up a MWM}

The characterization of an optimal solution given in section 3 provides useful insights for how to setup a MWM. In particular, the hopper weights should generate several different combinations of sums of weights whose densities cluster around the target package weight ($T$) providing many feasible options to the built-in knapsack algorithm to choose from (characteristic \# 2). In order to achieve this, the setting up of the hopper weights should form a very specific ``ladder" of weights $\{\mu_i\}$ whose values can be obtained with the heuristic in section 4 and the accompanying R program (see Appendix B). One observation of the actual solutions $\{\mu_i\}$ for the cases listed on Table \ref{tab5} is the following. As the maximum number of hoppers that can remain shut in a cycle increases, the best solutions call for hopper weight setpoints that are larger than if fewer hoppers could remain shut. The reason is that there is a need to fill the package with fewer opened hoppers, so each hopper should have more product. For instance, consider the solutions for $H=6$ and $H=7$ hoppers when up to to 2 or 3 of them can remain shut in a cycle (i.e., 6(2), 6(3), 7(2) and 7(3)). The hopper weights values $\{\mu_i\}_{i=1}^H$  found are:

\begin{tabular}{ccccccc}
6(2): & 149.9&149.2&133.7& 127.6& 108.2& 69.7\\
6(3):& 199.6 & 186.7&171.2&148.3&136.4&52.6
\end{tabular}

\noindent and

\begin{tabular}{cccccccc}
7(2): & 145.9&140.9&106.2&101.6&94.4&81.5&77.1\\
7(3):& 173.7&154.3&142.6&129.8&120.7&89.1&60.0
\end{tabular}

\noindent Notice how at the same time, there is also one hopper weight that should contain considerable less product, to help complete a package weight closer to the target a higher proportion of the times.

The most notable conclusion from our numerical experiments in Table \ref{tab5} is how adding more combinations by increasing the maximum number of hoppers shut (from up to two to up to three shut) but keeping the number of hoppers constant has a drastic decreasing effect in the average MSE, e.g., from 80.5 to 23.5 ($H=8$), from 111.9 to 8.98 ($H=9$), and from 12.8 to 1.23 ($H=12$). {\em For most practical purposes, the MSE's obtained (one gram of product) in the types of products packed with MWMs (with no package underweighted) implies that the solutions obtained with the proposed heuristic method are ``optimal enough" in practice.} These solutions (and the corresponding MSE analyses) can also be used to justify the acquisition of MWMs with large numbers of hoppers.

\section{Conclusions and further research}
In this paper, the setup problem of a multihead weigher machine has been studied. The hopper weight settings provided by the proposed optimization approaches will result in substantial savings over standard ad-hoc setup procedures used for companies utilizing MWMs. 

 A heuristic optimization model was developed based on a detailed characterization of what constitutes a near optimal solution to the MWM setup problem. The heuristic requires computations of moments of order statistics of correlated variables, and this becomes computationally intractable even for moderate size problems.
Using a lower bound approximation of the moments of smallest order statistics proved to be considerably faster. This lower bound heuristic is applicable for MWMs with several hoppers, considering up to 3 hoppers shut in a cycle. 

The behavior of the optimal solutions for MWM's with different number of hoppers indicate how the optimal setpoint weights decrease as the number of hoppers increases, but increase for most hoppers as the maximum number of hoppers allowed to remain shut increaases. It was shown how the mean square error of the packed weights decreases drastically as the number of combinations of hoppers shut or open increases, while keeping the number of hoppers fixed. Furthermore, great accuracy in packaging with minimum product ``given away" can be obtained with an optimally setup MWM with a large  number of hoppers ($H>10$) if only combinations with up to 3 hoppers remaining shut are considered. The analysis presented in this paper may be used to justify the adoption of advanced MWMs with several hoppers currently available in the market. An implementation of the proposed lower bound heuristic has been written in the R language (see Appendix B and supplementary material) and could be used for these purposes.

Further research can be directed to study other types of MWMs with more complex architecture and to study optimal setup problems of MWMs under objective functions different than the mean square error criterion investigated herein. For more complex MWMs with a very large number of hoppers $H$, or that mix several types of product in the same package, a simulation-optimization approach may be necessary to find its optimal settings.  Such an approach can also be useful to include economic considerations assigned to overfilled packages (rather than the MSE objective used here), as common in the canning literature (e.g., \cite{Arcelus,Pollock}).

\section*{Appendix A. Exact computations for the moments of the minimum of $K$ unconstrained, correlated normals}

Let $\bm X_{[1]}=\min (X_1,X_2,...,X_K)= \min(\bm X)$. Theorem 1 below provides expressions for E$(X_{[1]})$ and Var$(X_{[1]})$ when $\bm X$ is a multivariate normal with arbitrary mean and covariance matrix, not subject to any constraint. Afonja (1972) provided expressions for the computation of the moments of the maximum order statistics, which we modify in the Theorem below for the moments of the minimum order statistic.  In the theorem, $\phi_K(\bm x;\bm \Theta, \bm \Sigma)$ denotes the (multivariate normal) pdf of $\bm X$ with $\bm \Theta=(\theta_1,...,\theta_K)'$, $\bm \Sigma = \{ \sigma_{ij} \}$, and $\phi_K(\bm Z; \bm R)$ denotes the pdf of a standard multivariate normal with correlation matrix $\bm R$.

\begin{theorem}.  Let $\bm X \sim N(\bm \mu, \bm \Sigma)$ be a $K$-dimensional normal random variable. The $r^{th}$ moment of $X_{[1]}$ about the origin is given by
\[
m'_r(X_{[1]})=\sum_{i=1}^K \sum_{j=0}^r \left(\begin{array}{c} r\\j\end{array}\right) \theta_i^{r-j} \sigma_i^j \; m_j(Z_i)^{\bm b_i}_{{ \bm  R}_i}
\]
where $m_j(Z_i)$ denotes the marginal $j^{th}$ moment of a truncated standardized multivariate normal which is given by
\[ m_j(Z_i)^{\bm b_i}_{{ \bm  R}_i} = \int_{-\infty}^{b_{i1}}\int_{-\infty}^{b_{i2}} \cdots \int_{-\infty}^{b_{iK}} \; Z_i^j \phi_K(\bm Z; \bm R_i) \; d\bm Z
\]
with the upper limits of integration $\bm b_i'=(b_{i1},...,b_{iK})$ equal to:
\be
\label{b}
b_{ij} = \left\{ \begin{array}{lc}
\frac{\theta_j-\theta_i}{\sqrt{\sigma_{ii}+\sigma_{jj}-2 \sigma_{ij}}}, & j \neq i\\
{\small \infty}, & j = i
\end{array}
\right.
\ee
and the correlation matrix $\bm R_i$ is given by:
\be
\label{Ri}
\bm R_i = \left\{ \frac{\sigma_{ii}-\sigma_{ij'}-\sigma_{ij}+\sigma_{jj'}}{\sqrt{\sigma_{ii}+\sigma_{jj}-2 \sigma_{ij}} \sqrt{\sigma_{ii}+\sigma_{j'j'} - 2 \sigma_{ij'}}}\equiv r_{i,jj'}\right\}_{j' \neq i, \; \; j \neq i}
\ee
\end{theorem}
{\bf Proof.} If $X_{[1]}=\min(\bm X)$ then
\[
\mbox{E}(X_{[1]}^r) \equiv m_r'(X_{[1]}) = \int_{-\infty}^{\infty} \cdots \int_{-\infty}^{\infty} \min(\bm X)^r \phi_K(\bm X; \bm \Theta, \bm \Sigma) \; d\bm X
\]
where $X_{[1]}=X_i$ in the region $A_i= \left\{ \bm X : X_i< X_j ;\ \forall j \neq i, -\infty < X_i < \infty  \right\}$ with $\bigcup_{i=1}^K A_i = \mathbb{E}^K$ and $A_i \cap A_j = \varnothing (i \neq j)$, except for sets of zero measure where $X_i=X_j$ for some $i,j$. Note how regions $A_i$ correspond, for $K=2$, to areas number 1 and 4 in Figure \ref{fig1} for the unconstrained case, i.e., when $T=0$.  Therefore,
\begin{eqnarray*}
\mbox{E}(X_{[1]})^r &=& \int_{\bigcup_{i=1}^K A_i } \min(\bm X)^r \phi_K(\bm X; \bm \Theta, \bm \Sigma) \; d\bm X\\
&=&\int_{A_1} X_1^r \phi_K(\bm X; \bm \Theta, \bm \Sigma) \; d\bm X +\int_{A_2} X_2^r \phi_K(\bm X; \bm \Theta, \bm \Sigma) \; d\bm X + \cdots +\int_{A_K} X_K^r \phi_K(\bm X; \bm \Theta, \bm \Sigma) \; d\bm X\\
&=& \sum_{i=1}^K \int_{A_i} X_i^r \; \phi_K(\bm X; \bm \Theta, \bm \Sigma) \; d\bm X\\
&=& \sum_{i=1}^K \int_{A_i} (\theta_i + \sigma_i Z_i)^r \; \phi_K(\bm Z;  \bm R) \; d\bm Z
\end{eqnarray*}
where the last equality follows from $Z_i = (X_i-\theta_i)/\sigma_{i}$ $(\sigma_{i} \equiv \sqrt{\sigma_{ii}})$, $i=1,...,K$.

Following Wang and Mazumder (2005), we transform the integration region by redefining the $Z_j$ variables for $j \neq i$ according to
\begin{small}
\[
A_i= \left\{ \bm Z: -\infty < Z_j \equiv \frac{X_i - X_j-(\theta_i-\theta_j)}{\sqrt{\mbox{Var}(X_i-X_j)}} < \frac{\theta_j-\theta_i}{\sqrt{\mbox{Var}(X_i-X_j)}} \equiv b_{ij} ;\ \forall j \neq i, -\infty < Z_i \equiv \frac{X_i-\theta_i}{\sigma_i} < \infty  \right\}
\]
\end{small}
and form the $K \times 1$ vector
\be
\label{Zvector}
\bm Z = \left(\begin{array}{c}
Z_i \equiv \frac{X_i-\theta_i}{\sigma_i}\\[0.2cm]
\hline\\[-0.1cm]
Z_1 \equiv \frac{X_i - X_1-(\theta_i-\theta_1)}{\sqrt{\mbox{\scriptsize Var}(X_i-X_1)}}\\[0.3cm]
Z_2 \equiv \frac{X_i - X_2-(\theta_i-\theta_2)}{\sqrt{\mbox{\scriptsize Var}(X_i-X_2)}}\\[0.2cm]
\vdots\\[0.2cm]
Z_K \equiv \frac{X_i - X_K-(\theta_i-\theta_K)}{\sqrt{\mbox{\scriptsize Var}(X_i-X_K)}}
\end{array}
\right)
\ee
where the 2nd to $K$th elements below the line do not include $Z_i$. Let $\bm R_i$ the $K \times K$ covariance matrix of $\bm Z$, which, from the definition of $\bm Z$ (\ref{Zvector})  equals to:
\[
\bm R_i = \{\mbox{Corr}(X_i-X_j,X_i-X_j')\}_{ j\neq i, j' \neq i}
\]
which has entries as in (\ref{Ri}).
We then have:
\[
\mbox{E}(X_{[1]}^r) = \sum_{i=1}^K \int_{-\infty}^{\bm b_i}(\theta_i + \sigma_i \; Z_i)^r \; \phi_K(\bm Z;  \bm R_i) \; d\bm Z
\]
where $\bm b_i'=(b_{i1},b_{i2},...,b_{iK})$ ($b_{ii}=\infty$) as in (\ref{b}). Evaluating the binomial term inside the integral we finally get:
\be
\label{m'}
\mbox{E}(X_{[1]}^r) \equiv m_r'(X_{[1]}^r) = \sum_{i=1}^K \sum_{j=0}^r \left(\begin{array}{c} r\\j\end{array}\right) \theta_i^{r-j} \sigma_i^j \; m_j(Z_i)^{\bm b_i}_{{ \bm  R}_i}
\ee
where the moments of a truncated, standard multivariate normal are
\be
\label{mj}
m_j(Z_i)^{\bm b_i}_{{ \bm  R}_i} = \int_{-\infty}^{\bm b_i} Z_i^j \; \phi_K(\bm Z;  \bm R_i) \; d\bm Z \; \; \; \; \; \; \; \; \; \; \; \; \; \; \; \; \; \; \; \; \mbox{QED.}
\ee

\subsection*{Computation details for the first two moments of the smallest order statistic}

Formula (\ref{m'}) in Theorem 1 gives for $r=1$:
\be
\label{mean}
E(X_{[1]})=E[\min(\bm X)]=\sum_{i=1}^K \left( \theta_i \; m_0(Z_i)^{\bm b_i}_{{ \bm  R}_i}  + \sigma_i \; m_1(Z_i)^{\bm b_i}_{{ \bm  R}_i} \right)
\ee
and for $r=2$:
\begin{eqnarray}
\mbox{Var}(X_{[1]})&=&\mbox{Var}[\min(\bm X)]=E(X_{[1]}^2) - E(X_{[1]})^2 \nonumber\\
&=& \sum_{i=1}^K \left(\sigma_i^2 \; m_2(Z_i)^{\bm b_i}_{{ \bm  R}_i} + 2 \theta_i \; \sigma_2 \; m_1(Z_i)^{\bm b_i}_{{ \bm  R}_i} + \theta_i^2 \; m_0(Z_i)^{\bm b_i}_{{ \bm  R}_i}\right) - E(X_{[1]})^2
\label{var}
\end{eqnarray}

Tallis \cite{Tallis1,Tallis2} provides algorithms for computing the moments of a truncated multivariate normal (implemented in R package {\tt tmvtnorm}) which can in principle be used to compute (\ref{mj}) for $j=0,1,2$ and hence we would only compute (\ref{mean}) and (\ref{var}) and be done, as suggested by Afonja \cite{Afonja}. However, this is too slow, and a better approach is to use a recursive result from Wang and Mazumder \cite{WangMazumder}. These authors perform the ``trick" of putting the ith element of a vector in row 1 (as done in (\ref{Zvector}) above) which simplifies the handling of subindices in the computations required. The formulae by these authors have some  typos that we correct below.

For any upper bounds $\bm B$  and any standard normal random vector $\bm Y$ with correlation matrix $\bm V$:
\be
\label{m0}
m_0(Y_1)^{\bm B}_{ \bm  V} = \int_{-\infty}^{\infty} \underbrace{\int_{-\infty}^{b_1} \hdots \int_{-\infty}^{b_j} \int_{-\infty}^{b_p}}_{j \neq i} \phi_p(\bm Y, \bm V) d\bm Y
\ee
(assuming hereon that all vectors have their element $i$ put in element 1). Note that integration over $i$ is for all reals, so this is really an integral over $p-1$ dimensions).

In addition, if $\rho_{1,j} = \mbox{corr}(Y_1,Y_j) \; (j \neq i)$ and $\phi(b)$ is the standard normal density evaluated at $b$, Wang and Mazumder provide, for $r \in \{1,2\}$ the (here corrected) recursive expression:
\begin{small}
\be
\label{mr}
m_r(Y_1)^{\bm B}_{\bm  V} =(r-1) m_{r-2}(Y_1)^{\bm B}_{\bm  V} +\sum_{j=2}^K \left[
\rho_{1,j} \phi(b_j) \sum_{l=0}^{r-1} \left( \begin{array}{c} r-1\\ l \end{array} \right) \; (1-\rho_{1,j}^2)^{\frac{r-l-1}{2}} (-\rho_{1,j} b_j)^l m_{r-l-1}(Y_1)^{\bm B_j}_{\bm  V_j}  \right]
\ee
\end{small}
where
 \[
\bm B_j = \left\{ \bm Y_{(j)}: - \infty < Y_1 < \infty, - \infty \leq Y_l \leq \frac{b_l - b_j \rho_{l,j}}{\sqrt{1- \rho_{l,j}^2}}, l=2,3,...,K, l \neq j \right\}.
\]
We define a vector $\bm Y_{(j)}$, consisting of all elements in $\bm Y$ except of $Y_j$. Matrix $\bm V_j$ in (\ref{mr}) contains the partial correlations between $Y_m$ and $Y_n$ given $Y_j$, for any $Y_m$ and $Y_n$ in $\bm Y_{(j)}$ and is therefore of one dimension less than the covariance of the calling vector $\bm Y$:
\be
\bm V_j = \left\{ \frac{\rho_{m,n}-\rho_{j,m}\rho_{j,n}}{\sqrt{1-\rho_{j,m}^2} \sqrt{1-\rho_{j,n}^2}} \right\}_{m,n \neq j}
\label{Vj}
\ee
 where $\rho_{m,n} \equiv r_{i,mn}$ computed as in (\ref{Ri}) above.

Equation (\ref{mr}) results, for $r=1$ in:
\be
\label{m1}
m_1(Y_1)^{\bm B}_{\bm  V} = \sum_{l=2}^p \rho_{1,l} \phi(b_l) m_0(Y_1)^{\bm B_l}_{\bm V_l}
\ee
and for r=2 we get:
\be
\label{m2}
m_2(Y_1)^{\bm B}_{ \bm  V} = m_0(Y_i)^{\bm B}_{\bm  V} + \sum_{j=2}^p \rho_{1,j} \phi(b_{j}) \left[(1-\rho_{1,j}^2)^{1/2} \;
m_1(Y_1)^{\bm B_j}_{\bm  V_j} - (\rho_{1,j} \; b_{j}) \; m_0(Y_1)^{\bm B_j}_{\bm  V_j}  \right]
\ee

Using (\ref{m0}) and (\ref{m1}) we can obtain E$(X_{[1]})$ using (\ref{mean}).  Substituting  (\ref{m1}) and (\ref{m2}) into the variance formulae (\ref{var}) we get Var$(X_{[1]})$. However, the  computation of Var$(X_{[1]})$ needs to be done with care as the $m_r(Y)$ functions call each other recursively with  vector and matrix arguments of decreasing dimension.

Specifically, in the second call to $m_0(Y_1)$ we need to form the $(K-2) \times 1$ vector $\bm Y_{(j,l)}$, a vector equal to $\bm Y_{(j)}$ without element $l$. $\bm V_{j,l}$ is then the $(K-2) \times (K-2)$ partial correlation matrix between the elements of $\bm Y_{(j,l)}$ given $Y_l$. These partial correlations are obtained via (\ref{Vj}) but using the entries in matrix $\bm V_j$ rather than the entries in matrix $\bm R_i$. Likewise, $\bm B_{j,l}$ is obtained using the expressions for the bounds in $\bm B_j$, but using the correlations in $\bm V_j$ instead of those in $\bm R_i$.

We can now give a summary algorithm.
\begin{algorithm}[H]
\caption{$E(X_{[1]})$ and Var$(X_{[1]})$ computation}
\begin{algorithmic}[1]
\State Given:  $\bm X \sim N(\bm \Theta,\bm \Sigma)$, a $K$-dimensional vector with $\bm \Theta$ and $\bm \Sigma$ known.
\For{$i=1$ to K do}
\State $\bm Y=\bm Z_i$, $\bm B= \bm b_i$ using (\ref{b}), $\bm V = \bm R_i$ using (\ref{Ri}). This requires putting element $i$ in position 1, so $Z_1 \leftarrow Z_i$. This must be reflected in matrix $\bm R_i$.
\State Evaluate $m_0(Z_1)^{\bm B}_{\bm V}$ using (\ref{m0})
\State  Evaluate $m_1(Z_1)^{\bm B}_{\bm V}$ using (\ref{m1}). This is turns requires $m_0(Z_1)^{\bm B_j}_{\bm V_j}$
\State  Evaluate $m_2(Z_1)^{\bm B}_{\bm V}$ using (\ref{m2}). This is turns requires $m_1(Z_1)^{\bm B_j}_{\bm V_j}$ which in turn requires $m_0(Z_1)^{\bm B_{j,l}}_{\bm V_{j,l}}$
\State Accumulate sums over $i$ for E$(X_{[1]})$  (\ref{mean}) and Var$(X_{[1]})$ (\ref{var})
\EndFor
\State Return (\ref{mean}) and (\ref{var}), the first two moments of the smallest order statistics of the possibly correlated normal variables in $\bm X$.
\end{algorithmic}
\end{algorithm}


\section*{Appendix B. R software implementation of heuristic method}
The heuristic in section \ref{section3} was implemented in R (program {\tt OptimizeWeigher.R}). This program contains function {\tt constraintsAll} which evaluates the constraints (\ref{first})-(\ref{feasibility})  for given $\bm\mu, T,$ and $\alpha$. Function {\tt computeDet} evaluates the objective function (\ref{objective}). The program also allows the user to apply the modified lower bound heuristic which uses (\ref{LBmin}) in constraint (\ref{first}) --using the heuristic is actually the default.  The program also contains function {\tt computeMSD} (not used in the heuristic of section \ref{section3}) which evaluates expressions for $E[X_{[1]}]$, Var$[X_{[1]}]$,  $E[X_{[K]}]$ and Var$[X_{[K]}]$, the first two moments of the two extreme order statistics of a general $K$-dimensional normal distributed variable, a function that may be useful in other Applied Statistical problems.


\begin{thebibliography}{26}
\bibitem{Afonja}
Afonja, B., (1972), ``The Moments of the Maximum of Correlated Normal and t-Variates", {\em Journal of the Royal Statistical Society. Series B (Methodological)}, Vol. 34, No. 2, pp. 251-262.

\bibitem{Arcelus}
Arcelus, F.J., and Rahim, M.A., (1996), ``Reducing performance variation in the canning problem", {\em European Journal of Operational Research}, 94, pp. 477-487.

\bibitem{Beretta}
Beretta, A., and Semeraro, Q., (2012), ``On a RSM Approach to the Multihead Weigher Configuration", {\em Proceedings of the 11th Biennial Conference on Engineering Systems Design and Analysis, ASME 2012}, Vol. 1, pp. 225-233.

\bibitem{Bertsimas}
Bertsimas, D., Natarajan, K., and Chung-Piaw,  T., (2006), ``Tight Bounds on Expected Order Statistics", {\em Probability in the Engineering and Informational Sciences}, Vol. 20, No. 4, pp. 667-686.

\bibitem{Kameoka}
Kameoka, K.,  Nakatani, M., and  Inui, N., (2000). ``Phenomena in probability and statistics found
in a combinatorial weigher" (in Japanese). {\em Transactions of the Society of Instrument
and Control Engineers}, Vol. 36, pp. 388Ð394.

\bibitem{Karuno}
Karuno, Y., Nagamochi, H., and Wang, X., (2007). ``Bi-criteria food packing by dynamic programming", {\em J. of the Operations Research Society of Japan}, 50(4), pp. 376-389.

\bibitem{Karuno2}
Karuno, Y., Nagamochi, H., and Wang, X., (2010). ``Optimization problems and algorithms in double-layered food packing systems", {\em J. of Advanced Mechanical Design, Systems, and Manufacturing}, 4(3), pp. 605-615.

\bibitem{Imahori}
Imahori, S., Karuno, Y., Nagamochi, H., and Wang, X., (2011). ``Kansei engineering, humans and computers: efficient dynamic programming algorithms for combinatorial food packing problems", {\em, Int. J. of Biometrics}, 3(3), pp. 228-245.

\bibitem{Ishida}
Ishida Corporation Ltd., Japan, http://www.ishida.com/

\bibitem{Pollock}
Pollock, S., and Golhar, D., (1998). ``The canning problem revisited: the case of capacitated production and fixed demand", {\em European Journal of Operational Research}, 105, pp. 475-482.

\bibitem{Spall1987}
Spall, J.C., (1987). ``A stochastic approximation technique for generating maximum likelihood parameter estimates". {\em Proceedings of American Control Conference}, pp. 1161-1167.

\bibitem{Spall1992}
Spall, J.C., (1992). ``Multivariate Stochastic Approximation using a Simultaneous Perturbation Gradient Approximation". {\em IEEE Transaction on Automatic Control}, Vol. 37, pp. 332-341.

\bibitem{Spall1998a}
Spall, J.C., (1998). ``Implementation of the Simultaneous Perturbation Algorithm for Stochastic Optimization''. {\em IEEE Transaction on Aerospace and Electronic Systems}, Vol. 34, No. 3, pp. 817-823.

\bibitem{Tallis2}
Leppard, P., and Tallis, G.M., (1989), ``Algorithm AS 249: Evaluation of the Mean and Covariance of the Truncated Multinormal Distribution", {\em Journal of the Royal Statistical Society. Series C (Applied Statistics)}, Vol. 38, No. 3, pp. 543-553.

\bibitem{Prekopa}
Prekopa, A. (1995). {\em Stochastic Programming}, Kluwer Academic Publishers, Dordrecht, Boston.

\bibitem{Tallis1}
Tallis, G.M., (1965), ``Plane Truncation in Normal Populations", {\em Journal of the Royal Statistical Society. Series B (Methodological)}, Vol. 27, No. 2, pp. 301-307.


\bibitem{Wilhelm}
Wilhelm, S.,  (2013), ``Package ÔtmvtnormÕ", March 30, 2013, {\tt http://www.r-project.org/} downloaded April 2013.

\bibitem{WangMazumder}
Wang, B.,  and Mazumder, P., (2005), ``Multivariate Normal Distribution Based Statistical Timing Analysis Using Global Projection and Local Expansion", {\em Proceedings of the 18th International Conference on VLSI Design held jointly with 4th International Conference on Embedded Systems Design (VLSIDÕ05)},

\end{thebibliography}
\end{document}